\documentclass[aps,prd,a4paper,longbibliography,reprint,superscriptaddress,showpacs,floatfix,nofootinbib,amsmath,amssymb]{revtex4-1}
\usepackage{graphicx}
\usepackage[latin2]{inputenc}
\usepackage{dsfont}
\usepackage{mathrsfs}
\usepackage{slashed}
\usepackage{array}
\usepackage{colordvi}
\usepackage{rotfloat}
\usepackage{rotating}
\usepackage{color}
\usepackage[    bookmarks,
                 bookmarksopen = true,
                 bookmarksnumbered = true,
                 linktocpage,
                 colorlinks = true,
                 linkcolor = blue,
                 urlcolor  = blue,
                 citecolor = blue,
                 anchorcolor = green,
                 hyperindex = true,
                 hyperfigures]
		{hyperref}

\usepackage{cmap}

\newcommand{\dd}{\textmd{d}}
\newcommand{\be}{\begin{equation}}
\newcommand{\ee}{\end{equation}}

\newcommand{\Z}{\mathcal{Z}}
\newcommand{\A}{\mathcal{A}}
\renewcommand{\L}{\mathscr{L}}
\newcommand{\D}{\mathcal{D}}
\newcommand{\C}{\mathcal{C}}

\newcommand{\expv}[1]{\left \langle #1 \right \rangle}
\newcommand{\expvB}[1]{\Big \langle #1 \Big \rangle}
\newcommand{\tr}{\textmd{tr}\,}

\newcommand{\GeVt}{\,\textmd{GeV}^2}

\long\def\symbolfootnote[#1]#2{\begingroup%
\def\thefootnote{\fnsymbol{footnote}}\footnote[#1]{#2}\endgroup} 

\begin{document}
\title{Hadronic vacuum polarization and muon \boldmath{$g-2$} from magnetic susceptibilities on the lattice
}

\author{Gunnar~S.~Bali}
\affiliation{Institute for Theoretical Physics, Universit\"at Regensburg, D-93040 Regensburg, Germany}
\affiliation{Department of Theoretical Physics, Tata Institute of Fundamental Research, Homi Bhabha Road, Mumbai 400005, India.}
\author{Gergely~Endr\H{o}di}
\email[Corresponding author, email: ]{gergely.endrodi@physik.uni-r.de}
\affiliation{Institute for Theoretical Physics, Universit\"at Regensburg, D-93040 Regensburg, Germany}

\begin{abstract}
We present and test a new method to compute the hadronic vacuum polarization
function in lattice simulations. This can then be used, e.g.,
to determine the leading hadronic contribution to the anomalous
magnetic moment of the muon.
The method is based on computing susceptibilities with respect
to external electromagnetic plane wave fields and allows for a precision determination of  
both the connected and the disconnected contributions to the vacuum polarization.
We demonstrate that the statistical errors obtained with our method 
are much smaller than those quoted in previous lattice studies, 
primarily due to a very effective suppression of the errors of the disconnected 
terms. These turn out to vanish within small errors, enabling
us to quote an upper limit. We also comment on 
the accuracy of the vacuum polarization function determined from present experimental $R$-ratio data.
\end{abstract}


\maketitle

\section{Introduction}
\label{sec:intro}
The most precise measurement of the anomalous magnetic
moment of the muon, obtained by E821 at Brookhaven~\cite{Bennett:2006fi},
differs by more than three standard deviations from the
theoretical expectation.
At present, the uncertainties on the theory and on the experimental sides 
are of similar sizes.
For recent reviews and analyses,
see, e.g., refs.~\cite{Grange:2015fou,Blum:2013xva,Hagiwara:2011af,Davier:2010nc,Jegerlehner:2009ry}.
With the planned
E989 experiment at Fermilab~\cite{Grange:2015fou} and E34
at J-PARC~\cite{Aoki:2011cdr}, it is of utmost importance
to increase the precision of the standard
model prediction in line with the expected experimental improvement by a factor
of about five~\cite{Bennett:2006fi,Aoki:2011cdr,Grange:2015fou}. If
the discrepancy persisted at this even higher level of accuracy,
this should help
to pin down any particular beyond-the-standard-model
scenario and constrain the parameters of new interactions.
With an impressive QED five-loop evaluation~\cite{Aoyama:2012wk}
available, the theoretical uncertainty is
dominated by non-perturbative effects and, in particular,
by the leading hadronic contribution to the electromagnetic vacuum
polarization tensor, with the second biggest source of
uncertainty being the hadronic light-by-light scattering contribution.
The hadronic contribution to the
vacuum polarization tensor is also important in view of the running of the
electromagnetic fine structure constant and of the Weinberg weak mixing
angle~\cite{Moch:2014tta,Benayoun:2014tra,Bodenstein:2012pw,Hagiwara:2011af,Jegerlehner:2011mw}
from low to high scales.

The standard method~\cite{Blum:2002ii,Gockeler:2003cw}
employed in lattice calculations
of the leading hadronic
contribution to the anomalous magnetic moment
$a_{\ell}=(g_{\ell}-2)/2$ of a charged
lepton $\ell\in\{e,\mu,\tau\}$
consists of computing the renormalized vacuum polarization function
and inserting this into the leading-order QED formula.
The hadronic vacuum polarization tensor, which is the main
object of this study, is defined as
\begin{align}
\label{eq:vacu}
\Pi_{\mu\nu}(p) 
&= \int \!\dd^4\! x \, e^{ipx}  \expv{j_\mu(x) j_\nu(0)}\\\nonumber
&=\left(p_{\mu}p_{\nu}-\delta_{\mu\nu}p^2\right)\Pi(p^2)\,,
\end{align}
where
\begin{equation}
\label{eq:current}
j_{\mu}=\frac{q_u}{e}\bar{u}\gamma_{\mu}u
+\frac{q_d}{e}\bar{d}\gamma_{\mu}d+\frac{q_s}{e}\bar{s}\gamma_{\mu}s+\cdots
\end{equation}
denotes the quark electromagnetic current in position space
and $q_u/e=2/3$, $q_d/e=q_s/e=-1/3$ are the fractional
quark charges.
Due to electromagnetic current conservation,
$\Pi_{\mu\nu}$ is transverse and can be parameterized in terms
of a single vacuum polarization function $\Pi(p^2)$, where we employ
Euclidean spacetime conventions, i.e.\ the spacelike $p^2>0$
correspond to virtualities.
$\Pi(p^2)$ undergoes additive renormalization but the renormalized
combination
\begin{equation}
\label{eq:pren}
\Pi_{\mathrm{R}}(p^2)=\Pi(p^2)-\Pi(0)
\end{equation}
is ultraviolet finite.

It turns out that the leading hadronic contribution $a_\mu^{\mathrm{had,LO}}$ 
to the anomalous magnetic moment of the muon (see the
definition equation~(\ref{eq:amu}) below)
depends most strongly on $\Pi_{\mathrm{R}}(p^2)$ at relatively
small argument values.
Since small momenta correspond to large Euclidean distances,
naively implementing equation~(\ref{eq:vacu})
results in a bad signal over noise ratio in this region.
This becomes even worse for calculations
of the quark-line disconnected contributions,
which therefore have been neglected in almost all previous lattice studies.
Where these were taken into
account~\cite{Feng:2011zk,Francis:2014hoa,Burger:2015oya},
they dominated the statistical error.
Another problem of many past lattice attempts is a conceptual one:
$\Pi(0)$ often is extrapolated from $\Pi(p^2)$ at $p^2>0$ and
the parametrization used constitutes a source of
systematic uncertainty that
is difficult to estimate.

Here we propose methods that address both of the above issues.
The vacuum polarization at $p^2=0$ is shown to be equal to the
bare magnetic susceptibility of the system, which can 
be determined independently on the lattice. We investigate
different methods to achieve this, giving consistent results.
We also discuss how this quantity diverges as a function of
the lattice spacing towards the continuum limit.

Most importantly, we introduce a new method for computing
both the connected and the disconnected contributions
to the hadronic vacuum polarization function with unprecedented precision,
in particular at small momenta.
This consists of calculating $\Pi(p^2)$ at $p^2>0$
through the response of the system to oscillatory 
background electromagnetic fields. The new method is similar in spirit
to employing momentum sources~\cite{Martinelli:1994ty,Gockeler:1998ye},
allowing us to spend more effort on the low-$p^2$ points,
thereby increasing their precision, without wasting
resources on large momenta where $\Pi_{\mathrm{R}}(p^2)$
can easily be obtained within small relative errors, with a much smaller
impact on the predicted value of $a_{\mu}^{\mathrm{had,LO}}$.

The methods are tested on $N_f=2+1$ staggered ensembles at the physical point,
neglecting QED effects on the quark propagation which are
of a higher order in the fine-structure constant $\alpha$. In this situation,
due to $\sum_{f\in\{u,d,s\}}q_f=0$, disconnected
contributions vanish for $m_s=m_{ud}$ but need to be
taken into account for $m_s>m_{ud}$, which we do.
Since we neglect charm quark effects,
we have to restrict ourselves to $p^2< m_c^2$.
At high momenta our results can, however, be combined with measurements
of the $R$-ratio as well as with perturbation theory:
the non-singlet and singlet QCD contributions to the Adler function
have been calculated in massless QCD
to $\mathcal{O}(\alpha_s^4)$
in the strong coupling constant in refs.~\cite{Baikov:2010je} and
\cite{Baikov:2012zn}, respectively.

This article is organized as follows. In section~\ref{sec:review}
we review previous calculational strategies, followed by 
section~\ref{sec:new}, where 
we introduce our background field method and link this to
magnetic susceptibilities. We also discuss renormalization issues
and comment on relations between the Adler function
and the entropy density at high temperatures.
Finally, in section~\ref{sec:results}
we present the simulation setup
and first results, before we conclude. 
The equivalence between magnetic susceptibilities and the vacuum 
polarization is demonstrated in appendix~\ref{sec:appA}, and the 
details of our numerical implementation are discussed in appendices~\ref{appB} 
and~\ref{sec:appC}.

\section{Summary of previously employed methods}
\label{sec:review}
The leading hadronic contribution to the anomalous magnetic moment
is given as~\cite{Blum:2002ii,Lautrup:1971jf}
\be
a_{\ell}^{\mathrm{had,LO}} = 4\alpha^2\!\int_0^\infty \!\!\dd p^2 K_{\ell}(p^2)\Pi_{\mathrm{R}}(p^2)\,,
\label{eq:amu}
\ee
where the perturbative kernel function is defined as
\be
K_{\ell}(p^2) = \frac{m_{\ell}^2 \,p^2 Z_{\ell}(p^2)^3 \left[ 1-p^2Z_{\ell}(p^2)\right]}{1+m_{\ell}^2\,p^2Z(p^2)^2}
\ee
with
\be
Z_{\ell}(p^2) = \frac{\sqrt{1+4m_{\ell}^2/p^2}-1}{2m_{\ell}^2}\,.
\label{eq:amu2}
\ee
The renormalized hadronic vacuum polarization function is defined
in equations~(\ref{eq:vacu}) and (\ref{eq:pren}) above.
Note that the above expressions are valid to
leading-order in terms of the
QED fine-structure constant $\alpha=e^2/(4\pi)\approx 1/137$,
i.e.\ to $\mathcal{O}(\alpha^2)$,
which, at this order, can be pulled out of the integral.

In the limit of small momenta, where $\Pi_{\mathrm{R}}(p^2)\propto p^2$,
the argument of the integral has its maximum at
$p_0^2\approx (\sqrt{5}-2)m_{\ell}^2$. For the muon with
$m_{\mu}\approx 0.105\,$GeV
this implies $p_0^2\approx 0.0026\,\textmd{GeV}^2$:
an enormous volume would be necessary to resolve
this momentum region, at least without the use of twisted boundary
conditions~\cite{Sachrajda:2004mi,DellaMorte:2011aa},
since $\pi/p_0\approx 2\pi/m_{\mu}\approx 12\,\textmd{fm}$. 
Fortunately,
the integral as a whole turns out to be
dominated by somewhat higher momenta:
it still picks up about $50\%$ of its value from momenta
larger than $10\,p_0^2$. 
The predicted value of $a_{\mu}^{\mathrm{had,LO}}$ strongly
depends on $\Pi_{\mathrm{R}}(p^2)$ at these
still relatively small momenta $p^2\sim 0.03\,\textmd{GeV}^2$.
This is nicely illustrated, e.g., in ref.~\cite{Golterman:2014ksa},
in figure 3 of
ref.~\cite{DellaMorte:2011aa} and in figure 1 of ref.~\cite{Burger:2015oya}.

\subsection{Information from experiment}
The hadronic polarization tensor (and also the leading hadronic
contribution to the lepton anomalous magnetic moments~\cite{Lautrup:1971jf})
can be obtained by analytic continuation of the $R$-ratio
of the total cross section $\sigma(e^+e^-\rightarrow\mathrm{hadrons})$
over the tree-level
$e^+e^-\rightarrow\mu^+\mu^-$ expectation
(see, e.g., refs.~\cite{Adler:1974gd,Bernecker:2011gh}):
\be
\label{eq:rrat}
\Pi_{\mathrm{R}}(p^2) = \frac{p^2}{12\pi^2} \int_{4m_{\pi}^2}^\infty\!\! \dd s\,
\frac{R(s)}{s(s+p^2)}\,.
\ee
$R$-ratio measurements~\cite{Davier:2010nc,Hagiwara:2011af} can in principle
be augmented by other experimental data,
including $\tau$-decays into
final states containing $\pi^+\pi^-$, see, e.g., ref.~\cite{Benayoun:2012wc}.

\begin{figure}[ht!]
 \centering
 \includegraphics[width=8cm]{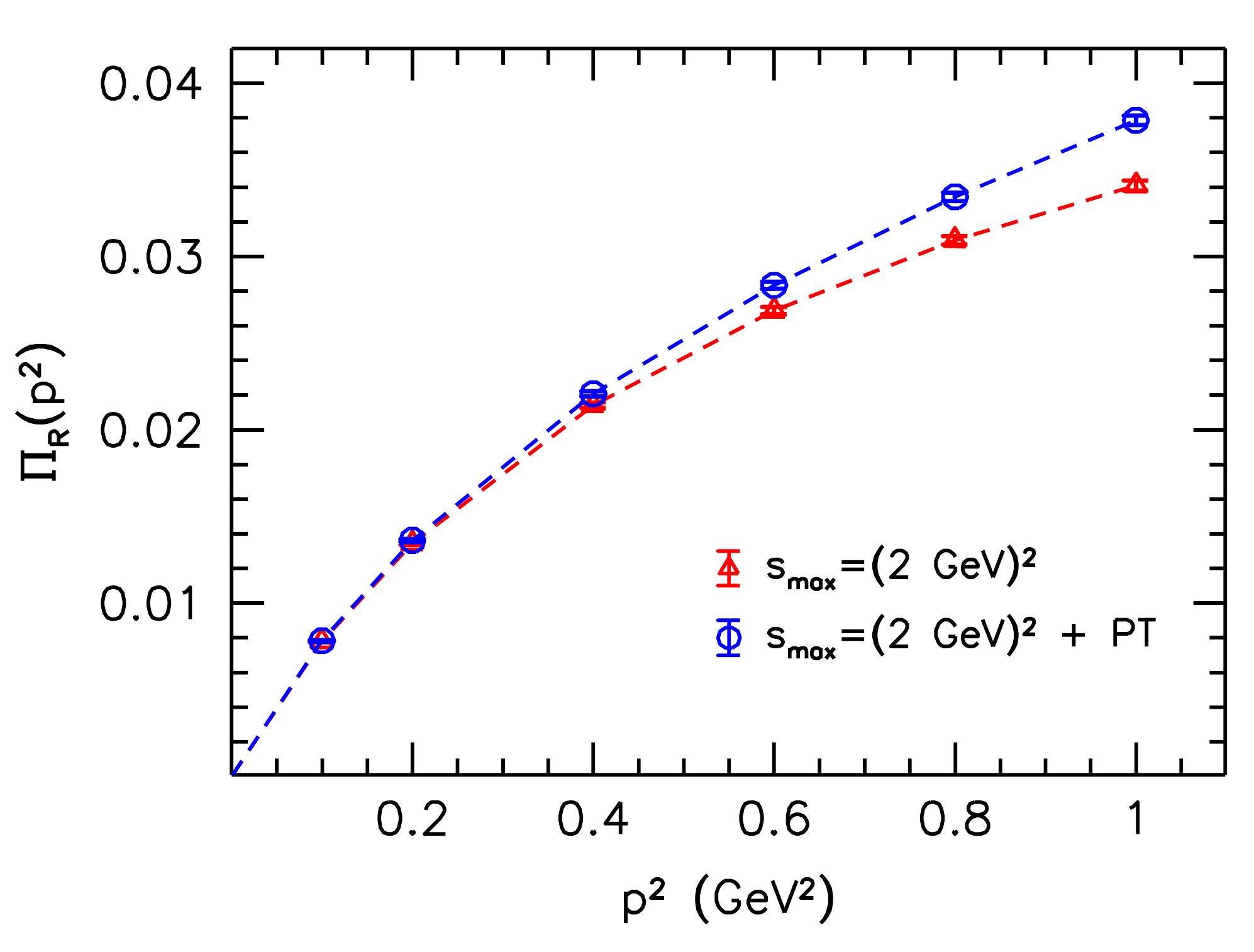}
 \caption{\label{fig:exp}
The renormalized vacuum polarization determined from the experimental
$R$-ratio~\protect\cite{Bogdan}, performing the integral~\protect(\ref{eq:rrat}) up
to $s=s_{\max}=(2\,\textmd{GeV})^2$, where three quark flavors are
active. Also indicated is the result of the integral
supplemented by three-flavor perturbation theory for $s>(2\,\textmd{GeV})^2$.
}
\end{figure}

In figure~\ref{fig:exp} we show the so-determined renormalized vacuum polarization
as a function of $p^2$~\cite{Bogdan}.
The present relative precision of $\Pi_{\mathrm{R}}$
is 0.64\% at $p^2=0.025\,\textmd{GeV}^2$,
increasing to 0.74\% at $p^2=0.6\,\textmd{GeV}^2$~\cite{Davier:2010nc,Bogdan}.
Achieving a statistical error below 1\% around $p^2=0.03\,\textmd{GeV}^2$ 
already constitutes an enormous challenge for present-day lattice determinations,
and such results still need to be extrapolated to the infinite volume
and continuum limits and, often, to physical quark masses.
In principle, lattice data at large $p^2$ values -- where discretization errors are 
enhanced -- can be substituted by results from the $R$-ratio.
Such a combined strategy may prove optimal for an accurate determination of 
$a_\mu^{\mathrm{had,LO}}$, once sufficiently precise lattice results
become available.

\subsection{Lattice determinations of \boldmath{$\Pi(0)$}}
In the past, two strategies have been used to obtain the zero-momentum subtraction
$\Pi(0)$. One possibility are fits of $\Pi(p^2)$ data, e.g., to pole
parametrizations, assuming
vector dominance~\cite{Gockeler:2003cw,Aubin:2006xv,Boyle:2011hu,DellaMorte:2011aa,Gregory:2013taa}, which
is also suggested to be the dominant contribution by
chiral perturbation theory~\cite{Aubin:2006xv}. Extending the
fit region towards large momenta,
such pole ans\"atze have
also been combined with 
polynomial parametrizations~\cite{Aubin:2006xv,Feng:2011zk,Burger:2013jya,Burger:2015oya}, motivated by perturbation theory.
Another popular and less model-dependent way to obtain
the normalization is through
Pad\'e approximants~\cite{DellaMorte:2011aa,Aubin:2012me,Golterman:2014ksa,Marinkovic:2015zaa}.

As an alternative, one can compute derivatives of
$\Pi_{\mu\nu}(p)$ from its definition in terms of
the continuum Fourier transformation~(\ref{eq:vacu}).
Then the divergent contribution that
needs to be subtracted from $\Pi(p^2)$ can, e.g., be obtained
via
\begin{align}
\Pi(0)&=\left.-\frac{1}{2}\frac{\partial^2}{\partial p_{\mu}^2}\Pi_{\nu\nu}(p)\right|_{p=0} \qquad\quad(\mu\neq\nu)\nonumber\\\label{eq:tmoment}
&=\frac{1}{2}\int\!\dd^4\!x\,x_{\mu}^2\,\langle j_{\nu}(x)j_{\nu}(0)\rangle=\frac{1}{2}\int\!\dd t\,t^2\,G(t)\,,
\end{align}
where no summation over $\nu$ is implied and 
in the last step we identified $\mu$ with the time-direction,
to emphasize the correspondence to the second $t$-moment of
a zero-momentum projected two-point function
\begin{equation}
\label{eq:get}
G(t)=\int\!\dd^3\!r\,
\langle j_{i}(\mathbf{r},t)j_{i}(0)\rangle\,.
\end{equation}
This method was used, e.g., in
refs.~\cite{Francis:2013fzp,Malak:2015sla}, to
obtain this subtraction.

Finally, in ref.~\cite{deDivitiis:2012vs} the expansion
of the two-point current-current correlation function in
powers of $p_{\mu}$ is carried out already on the level of
quark propagators.
This enables the direct computation of
$\Pi(0)=\partial^2\Pi_{12}/(\partial p_1\partial p_2)|_{p=0}$,
without relying on a continuum formula. However, this comes
at the price of computing the expectation value of an operator
involving up to four fermion matrix inversions, without even considering
disconnected contributions.

\subsection{Lattice determinations of \boldmath{$\Pi(p^2)$}, \boldmath{$\Pi_{\mathrm{R}}(p^2)$}
or moments thereof}
The lattice vector Ward-Takahashi identity
reads $\hat{p}_{\mu}\Pi_{\mu\nu}=0$ and
therefore~\cite{Gockeler:2000kj,Blum:2002ii,Gockeler:2003cw}
\begin{equation}
\Pi_{\mu\nu}(p^2)=\left(\hat{p}_{\mu}\hat{p}_{\nu}-\delta_{\mu\nu}\hat{p}^2\right)\Pi(p^2)\,,
\label{eq:latticeWI}
\end{equation}
where $\hat{p}_{\mu}=(2/a)\sin(ap_{\mu}/2)$. This change that
affects $\Pi(p^2)$ at high momenta
has been implemented in almost all
lattice studies, as well as
a modified phase $e^{ipx}\mapsto
e^{ip(x+a\hat{\mu}/2-a\hat{\nu}/2)}$ within the
Fourier sum for $\mu\neq\nu$.

Most lattice evaluations use what we will refer to below as 
the conventional method.
This amounts to directly computing the lattice version of equation~(\ref{eq:vacu}),
see, e.g., refs.~\cite{Blum:2002ii,Gockeler:2003cw,Aubin:2006xv,Feng:2011zk,Boyle:2011hu,Burger:2013jya,Gregory:2013taa,Malak:2015sla}.
In some cases, lower momenta
were made accessible by the use of
twisted boundary conditions~\cite{DellaMorte:2011aa,Aubin:2013daa,DellaMorte:2014rta}.
Very recently, another interesting method, stochastically
averaging over different twists, has been suggested~\cite{Lehner:2015bga}
that reduces finite volume effects and allows to realize
very small momenta. The main problem of modifying the fermionic
boundary conditions is that this cannot
easily be extended to incorporate quark-line disconnected contributions.

Obviously, equation~(\ref{eq:amu}) can be Taylor expanded in powers
of $p^2$ and the coefficients can be related to those
of the corresponding
expansion of $\Pi_{\mathrm{R}}(p^2)$. Generalizing
equation~(\ref{eq:tmoment}) above, the first and higher order derivatives
of $\Pi_{\mu\mu}$ with respect to $p^2$ can be obtained, computing
$t^2$-moments of two-point zero-momentum (spatial) projected current-current
correlators. This was explored within ref.~\cite{Feng:2013xsa}
and carried out for the first few moments of the
connected strange and charm quark
contributions to $a_{\mu}^{\mathrm{had,LO}}$ in ref.~\cite{Chakraborty:2014mwa}.
In ref.~\cite{Bernecker:2011gh}
the anomalous magnetic moment was directly related to the
zero momentum projected current-current two-point function.
This approach was then employed, e.g.,
in refs.~\cite{Francis:2013fzp,DellaMorte:2014rta,Francis:2014hoa}.

So far, disconnected contributions have been included in very few lattice 
studies~\cite{Feng:2011zk,Francis:2014hoa,Burger:2015oya}.
While their effect seems to be small, the associated
statistical error exceeds that of the connected terms.
Here we will find that this need not be the case.
There exist theoretical expectations
regarding the size of flavor singlet contributions:
exploiting the fact that
$m_{\omega}, m_{\phi}>m_{\rho}$, it was
demonstrated~\cite{Francis:2014hoa} that
the ratio of the disconnected contribution over the total
momentum-projected current-current two-point function
$G(t)$, defined in equation (\ref{eq:get}), approaches
the value $-1/9$, in the limit of large Euclidean times for
$N_f=2+1$ quark flavors. This ratio will, however, not automatically
propagate into $\Pi_{\mathrm{R}}(p^2)$ that depends on $G(t)$ at
all times $t$.
Next-to-leading order chiral perturbation
theory arguments show the disconnected contribution to
also account for $-1/9$ of the total
$\Pi_{\mathrm{R}}(p^2)$~\cite{DellaMorte:2010aq}.
However, this observation builds on the fact that the 
correlator of the iso-singlet vector current 
$\bar u \gamma_\mu u + \bar d \gamma_\mu d$ 
is momentum-independent to this order of chiral perturbation theory ---
which we found is not at all satisfied by the lattice data.
Thus, direct computation of the disconnected terms cannot be avoided 
in a systematic study.
Our numerical results will shed light onto the size 
of the disconnected contribution at low $p^2$.

\section{Vacuum polarization from susceptibilities}
\label{sec:new}
\subsection{The method}
The photon vacuum polarization tensor~(\ref{eq:vacu})
can also be interpreted as a momentum space current-current correlation function
\be
\Pi_{\mu\nu}(p)
=\frac{1}{V_4} \expv{\widetilde{j_\mu}(p) \widetilde{j_\nu}(-p)}, 
\label{eq:Piq}
\ee
where $V_4$ denotes the four-dimensional volume of the system
and $\widetilde{j_{\mu}}$ is the
Fourier transform of the electromagnetic current defined
in equation~(\ref{eq:current}):
\be
\widetilde{j_\mu}(p) = \int \dd^4\! x\, e^{ipx} j_{\mu}(x)\,.
\ee
Depending on the lattice definition of $j_{\mu}$, 
the polarization tensor~(\ref{eq:Piq})
may or may not renormalize multiplicatively with $Z_V^2$.
Here, we work with a conserved current, i.e.\ $Z_V=1$.

In the following we will relate the vacuum polarization
to the leading response 
of the free energy density $f$ of the system to background
electromagnetic fields. 
To illustrate the relation between the two objects on a qualitative level,
it is instructive 
to represent the vacuum polarization tensor by the diagram
\begin{figure}[h!]
\centering
\includegraphics[width=3cm]{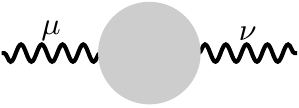}
\end{figure}

\noindent
where a momentum $p$ flows in and out of the photon legs. Here, the gray blob 
indicates all possible closed loops formed by quark and gluon propagators ---
i.e.\ the perturbative expression for the free energy density $f$. 
The legs may 
be thought of as photons corresponding to a background electromagnetic field
$A_\mu$
with momentum $p$.
Pulling out these legs is achieved by taking appropriate derivatives 
of $f$ with respect to the background field. 
While background electric fields turn the Euclidean QCD
action complex and are thus
problematic in lattice simulations, background magnetic fields
can be realized without complications. Employing the latter gives
access to the spatial components
$\Pi_{ij}$ and hence to all components $\Pi_{\mu\nu}$ since in Euclidean
spacetime at zero temperature the indices can be relabelled at will.

To find the background field corresponding to $\Pi_{\mu\nu}(p)$, we define 
the magnetic fields
\be
\mathbf{B}^{p}(x) = B\sin(px)\,\mathbf{e}_3\,, \qquad
\mathbf{B}^{0} = B\,\mathbf{e}_3\,,
\label{eq:Bfields}
\ee
pointing in the third spatial direction. 
While $\mathbf{B}^p$ is an oscillatory magnetic field with oscillation frequency $p$, 
$\mathbf{B}^0$ is a homogeneous background. 
The corresponding susceptibilities are obtained as the second 
derivatives of the free energy density with respect to the amplitude of 
the magnetic field:
\be
\chi_p = -\left.\frac{\partial^2 f[\mathbf{B}^{p}]}{\partial (eB)^2}\right|_{B=0}\,.
\label{eq:chidef1}
\ee
These susceptibilities are 
normalized by the square of the elementary charge $e>0$ to ensure that
only the renormalization group-invariant combination $eB$ appears in the definitions.
Note that $\chi_p$ can be evaluated on gauge ensembles generated at $B=0$.

The explicit calculation in appendix~\ref{sec:appA} shows that
\be
2\chi_p = \Pi(p^2)\,, \qquad\chi_0 = \Pi(0)\,.
\label{eq:mainresult}
\ee
These relations form a new representation of the vacuum polarization
function in terms  of susceptibilities with respect to the magnetic fields
defined in equation~(\ref{eq:Bfields}) 
and are the main result of this article.

Unlike the conventional method,
where the polarization function is
extracted from the same set of position space current-current
correlators for all momenta, equation~(\ref{eq:mainresult}) gives access 
to $\Pi(p^2)$ at one single lattice momentum $p$. While this
certainly increases the costs of calculating $\Pi$ over a large
range of momenta, it also allows for a better signal-to-noise ratio
within momentum regions of particular interest. 
As argued above, for the determination of the hadronic
contribution to the muon anomalous
magnetic moment $a_\mu^{\mathrm{had,LO}}$,
low momenta $p^2\sim 0.03\,\textmd{GeV}^2$
are  much more important than the high-$p$ region.
While $\expv{j_\mu(x)j_\nu(0)}$ mixes information about 
all allowed values of $p$, here such a mixing is avoided. 

Just as the vacuum polarization tensor, $\chi_p$ and $\chi_0$
can also be separated into
connected and disconnected contributions. We will demonstrate in
section~\ref{sec:results} below that,
using this new approach, an unprecedented accuracy can be achieved
for both the connected and the disconnected contributions to the
vacuum polarization function, already at moderate computational costs. 
An additional advantage of the method is that it gives direct
access to $\Pi(0)$.

To summarize, to arrive at a prediction for $a_\mu^{\mathrm{had,LO}}$ it
is desirable to
improve the accuracy in the low-$p$ region and to calculate $\Pi(0)$
independently. The method we propose
accomplishes both of these requirements.

\subsection{Renormalization}

Before presenting the details of the implementation and our numerical results,
it is instructive to discuss the renormalization properties 
of $\chi_0$ in more detail. 
Equation~(\ref{eq:mainresult}) reveals that 
the homogeneous susceptibility is additively divergent, just as $\Pi(0)$. 
To see where this divergence comes from, let us consider the multiplicative renormalization of the
background magnetic field (and the corresponding renormalization 
of the electric charge),
\be
e^2 = Z_e^{-1} e_r^2\,, \qquad B^2 = Z_e B_r^2\,, \qquad eB=e_rB_r\,,
\ee
with the renormalization factor
\be
Z_e = 1 + 2b_1 e_r^2 \log(\mu a)\,,
\ee
where $a$ is the lattice spacing (inverse of the regulator) and
$\mu$ the renormalization scale.
Notice that since the magnetic field is external and has no dynamics,
only the lowest-order 
QED $\beta$-function coefficient --- denoted as $b_1$ --- appears
in $Z_e$~\cite{Schwinger:1951nm,Dunne:2004nc,Bali:2014kia}. 

The total free energy density $f^{\mathrm{tot}}$ of the system is the sum of
$f$ and the energy $B^2/2$ of the magnetic field. 
Since varying the background field
should not change the ultraviolet 
properties of the system, $f^{\mathrm{tot}}$ must be free of $B$-dependent
divergences. 
This implies that the divergence of the pure magnetic energy
\be
\frac{B^2}{2} = \frac{B_r^2}{2} + b_1 (eB)^2 \log(\mu a) 
\ee
is exactly cancelled by an analogous divergence of $f$. 
Plugging this divergence into the definition~(\ref{eq:chidef1}),
we obtain
\be
\chi_0 = 2 b_1(a)\, \log(\mu a)\,.
\label{eq:chiren}
\ee
The renormalization scale $\mu$ is fixed by the requirement that there
should be no 
finite quadratic terms in $f^{\mathrm{tot}}$ other than $B_r^2/2$~\cite{Schwinger:1951nm}.
Let us emphasize that $b_1$ is the lowest-order coefficient of the QED $\beta$-function, however, 
with all QCD corrections taken into account. 
To highlight this, we explicitly indicate the dependence of
$b_1$ on the lattice spacing.
Perturbatively, this reads~\cite{Baikov:2012zm}
\be
b_1(a) = \sum_{f=u,d,s} (q_f/e)^2 \frac{1}{4\pi^2} \left[ 1 + \frac{g^2(a)}{4\pi^2} + \ldots \right]\,,
\label{eq:b1p}
\ee
where  $g^2(a)$ is the QCD coupling.
Equation~(\ref{eq:chiren}) allows to connect lattice results for $\chi_0$ 
to perturbation theory, once the lattice spacing is small enough,
cf.\ ref.~\cite{Bali:2014kia}.

\subsection{Implication for hot or dense QCD}

As a side-remark, we mention that the
correspondence~(\ref{eq:mainresult}) can be generalized 
to high temperatures.
In this case
it results in a relation between the entropy density and the perturbative 
Adler function~\cite{Bali:2014kia}. The latter is defined as
the logarithmic derivative of the polarization function
with respect to the squared momentum~\cite{Adler:1974gd}:
\be
D(p^2) = 12\pi^2 \frac{\partial \Pi(p^2)}{\partial \log(p^2)}\,.
\label{eq:Adler}
\ee

Let us consider QCD at a high temperature $T$, which exceeds
all other dimensionful scales in the system.
In this limit, the argument of $\Pi$ is set by a thermal scale
$\mu_{\mathrm{th}}=2\pi T$, 
leading to the correspondence 
$\Pi(\mu_{\mathrm{th}}^2)\leftrightarrow \chi_0(T^2)$. 
(The susceptibility at 
high temperatures indeed only depends on $T^2$~\cite{Bali:2014kia}.)
For the Adler function, this implies the relation
\be
D(\mu_{\mathrm{th}}^2) \longleftrightarrow 12\pi^2 \frac{\partial \,\chi_0}{\partial \log T^2} =
6 \pi^2 T \left.\frac{\partial^2 s}{\partial (eB)^2}\right|_{B=0}\,,
\label{eq:drel}
\ee
where in the second step we used the definition of the entropy density 
$s\equiv -\partial f/\partial T$. 
Equation~(\ref{eq:drel}) reveals that the leading dependence of
the entropy density 
on the magnetic field at high temperatures is fixed by the Adler function,
i.e.\ by perturbative QED physics. 
Repeating the above argument with $T$ replaced by a chemical potential $\mu$ 
(or by an isospin chemical potential $\mu_{\mathrm{I}}$) gives an analogous relation 
for the quark number density $n=-\partial f/\partial \mu$
at high $\mu$ (or for the isospin density $n_{\mathrm{I}} = -\partial f/\partial \mu_{\mathrm{I}}$ at high $\mu_{\mathrm{I}}$).
We believe these are highly non-trivial findings.

\section{Simulation details and numerical results}
\label{sec:results}

We employ the $N_f=2+1$ staggered lattice ensembles~\cite{Bali:2011qj,Bali:2012zg}
generated at physical pion and kaon masses. 
Each ensemble --- summarized in table~\ref{tab:1} --- consists of a
hundred to a few hundred effectively statistically decorrelated
configurations.
Details of the simulation algorithm and of the lattice setup can be 
found in refs.~\cite{Aoki:2005vt,Borsanyi:2010cj,Bali:2011qj}.

\begin{table}[ht!]
 \caption{\label{tab:1}Lattice ensembles investigated; the largest lattice spacing reads $a_0=0.29\,\textmd{fm}$.}
 \begin{ruledtabular}
\begin{tabular}{ccccc}
 $N_s$ & $N_t$ & $\beta$ & $a\,[\textmd{fm}]$ & $\log(a/a_0)$ \\ \hline
 24 & 32 & 3.45 & 0.290 & 0 \\
 24 & 32 & 3.55 & 0.216 & -0.295 \\
 32 & 48 & 3.67 & 0.153 & -0.636 \\
 40 & 48 & 3.75 & 0.125 & -0.843 \\
 40 & 48 & 3.85 & 0.099 & -1.078 \\
 \end{tabular}
\end{ruledtabular}
\end{table}

\begin{figure}[ht!]
 \centering
 \includegraphics[width=8cm]{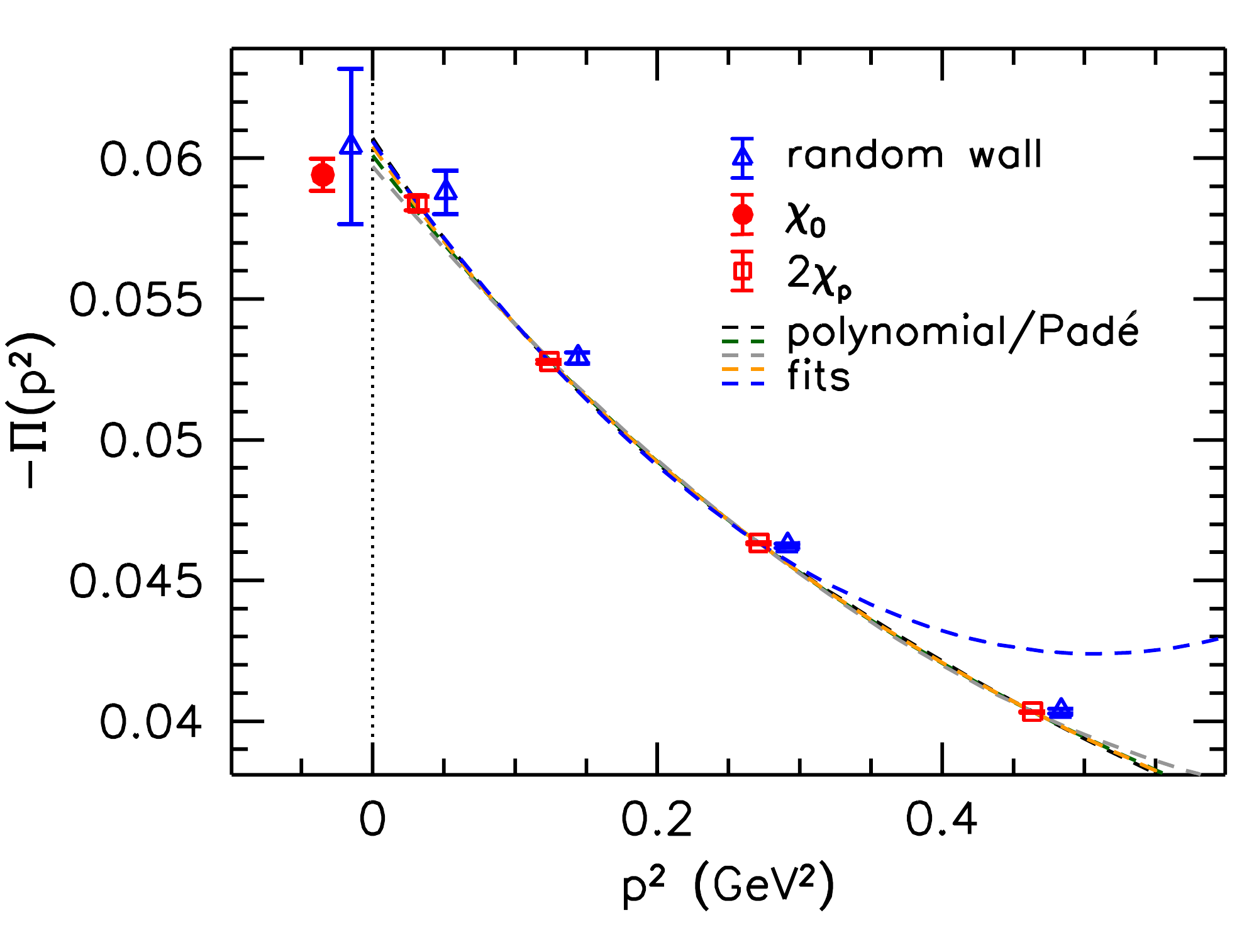}
 \caption{\label{fig:lowp}
The low-momentum region of the oscillatory susceptibilities as measured on the 
$24^3\times 32$ configurations at $\beta=3.45$. The curves
correspond to polynomial- and Pad\'e-type 
extrapolations of $2\chi_p$ to $p=0$. 
The direct determination $\chi_0$ is shifted horizontally to
the left for better visibility.
Also included are results obtained using random wall sources,
displaced horizontally to the right.
}
\end{figure}

\subsection{Oscillatory susceptibilities}

First we discuss results on the susceptibilities $\chi_p=\Pi(p^2)/2$ with respect to 
the oscillatory backgrounds. These are determined via the noisy estimator 
technique described in appendix~\ref{appB}.
A typical set of low-momentum results is shown in figure~\ref{fig:lowp}.
The data include both the connected and the disconnected contributions
to $\Pi(p^2)$.
The figure also includes results obtained via the conventional method, however, 
employing stochastic wall sources (for our 
numerical implementation, see appendix~\ref{sec:appC}). 
The comparison reveals full agreement between the 
two approaches. The statistical error of the random wall data 
increases towards small momenta, whereas
it remains tiny even for the lowest
non-vanishing $p^2$-value shown for the oscillatory susceptibilities. 
Note that the number of inversions employed to obtain the data point at the lowest momentum 
was the same, $N_{\mathrm{inv}}=3000$, for both approaches.

In most previous lattice studies, $\Pi(0)$ 
was obtained by extrapolating $\Pi(p^2)$ to zero. 
Some possible extrapolations, employing polynomials or Pad\'e approximants,
fitted over various ranges in $p^2$, are included in the figure. These fits 
are also compared to the direct determinations via the homogeneous 
susceptibility $\chi_0$ (see section~\ref{sec:homchi} below) and via 
the zero-momentum projected current-current correlation function $G(t)$
according to equation~(\ref{eq:tmoment}), again obtained using random wall sources.
Within
their scatter, at $p^2=0$ the extrapolations agree with the direct
determinations. 
We remark that increasing the precision for the lowest few momenta
stabilizes such extrapolations tremendously.

\subsection{Homogeneous susceptibility and renormalized vacuum polarization}
\label{sec:homchi}

The susceptibility $\chi_0$ with respect to a homogeneous background is of interest 
for QCD thermodynamics in magnetic fields and has been the subject of detailed 
studies in the past few years. 
The determination of $\chi_0$ is considerably more complicated than that of $\chi_p$ 
due to the quantization of the magnetic flux $\Phi$. 
On the one hand, oscillatory magnetic fields have zero flux 
and can be varied continuously, allowing 
for a direct differentiation with respect to $B$. On the other hand, 
homogeneous fields have nonzero flux. Therefore, such a differentiation
cannot be carried out to determine $\chi_0$,
see appendix~\ref{appB}.
Several approaches, summarized in refs.~\cite{Bali:2014kia,D'Elia:2015rwa},
have been developed recently to overcome this problem.
Here we compare results obtained using the finite difference
method~\cite{Bonati:2013lca}, 
the generalized integral method~\cite{Bali:2014kia} and 
the half-half method~\cite{Levkova:2013qda}.
The former two approaches are based on simulations at non-zero magnetic
flux values, numerically differentiating the results with respect to $\Phi$. 
The half-half method involves calculating expectation values directly at $B=0$,
employing a setup where the magnetic field is positive in one half 
and negative in the other half of the lattice.
In this case, since the total flux is zero, a direct differentiation 
with respect to the amplitude is possible.
However, the discontinuity of the magnetic field turns out to
dramatically enhance
finite volume effects in $\chi_0$, see 
below.\footnote{These finite volume effects cancel
to a large extent in the difference
$\chi_0(T)-\chi_0(T=0)$~\cite{Ludmilla}, which is relevant for
QCD thermodynamics in background magnetic fields.}

\begin{figure}[b]
 \centering
 \includegraphics[width=8cm]{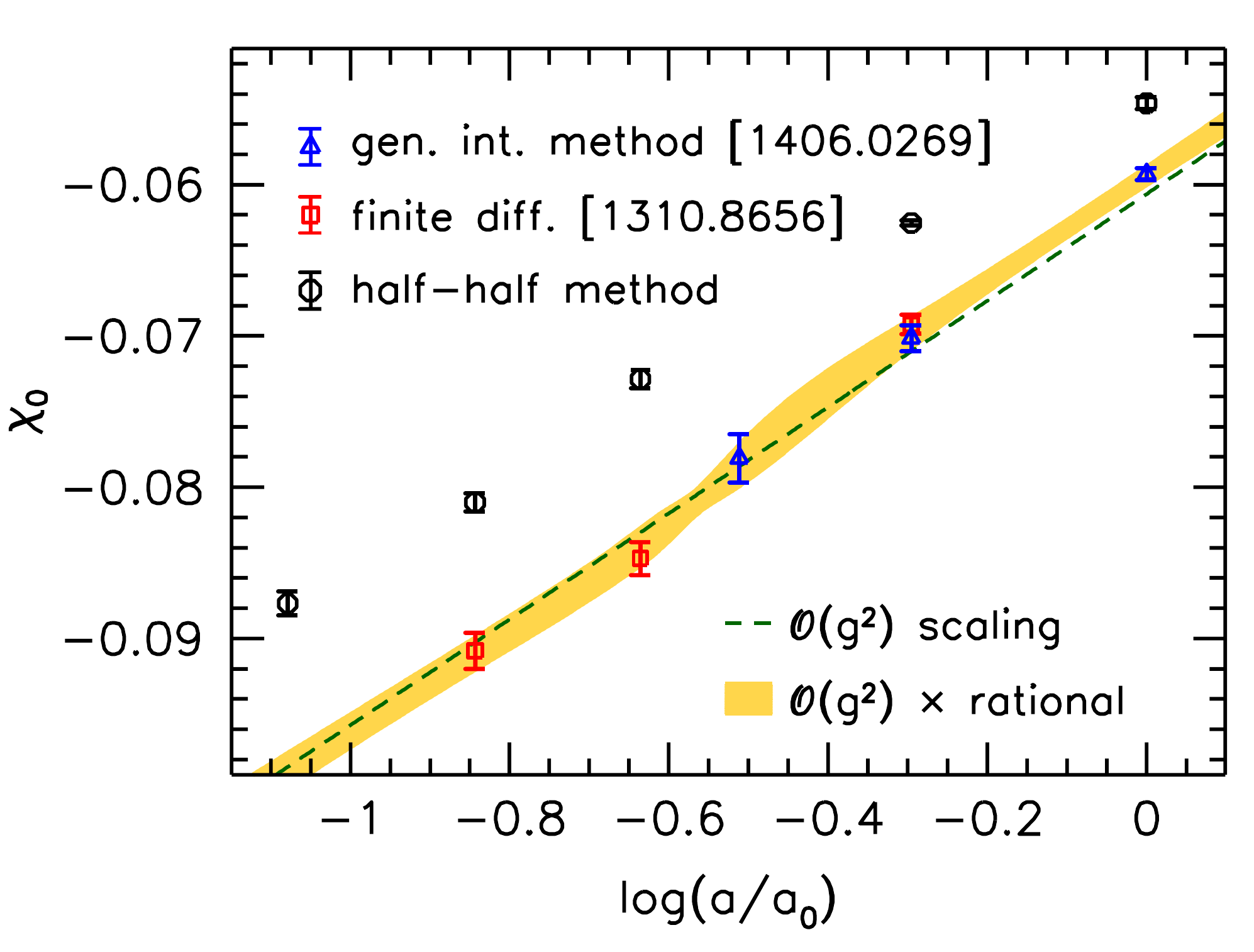}
 \caption{\label{fig:chihom}
 Magnetic susceptibility with respect to a homogeneous background as 
 a function of the logarithm of the lattice spacing
($a_0=0.29\,\textmd{fm}$), using three different approaches
(the generalized integral method~\cite{Bali:2014kia},
the finite difference method~\cite{Bonati:2013vba,Claudio}
and data generated in this study using the half-half method~\cite{Levkova:2013qda}).
Also included are a comparison to $\mathcal{O}(g^2)$ perturbation theory
and a parametrization via a rational ansatz.
}
\end{figure}

In figure~\ref{fig:chihom}, we compare all three approaches.
The results from the
generalized integral method and from the finite difference approach
are taken from refs.~\cite{Bonati:2013vba,Claudio} while
the half-half results are new. Not all lattice spacings
are covered by all the methods.
While the results of the generalized integral method\footnote{Here
we compare data obtained on $N_t>N_s$ zero-temperature lattices.
On the configurations of ref.~\cite{Bali:2014kia} at finite (but low)
temperatures, $\chi_0$ was found to have slightly 
smaller absolute values for fine lattices of table~\ref{tab:1} ($\beta\ge 3.67)$.}
and of the finite
difference approach are consistent with each other,
the half-half approach consistently underestimates the magnitude of the 
susceptibility. 
The difference between that approach on the one hand and the other two methods
on the other hand is found to be as large as $10\%$ 
and reduces only very slowly with increasing lattice volumes.\footnote{The
comparison between the half-half method and the generalized
integral method on our coarsest lattice, already 
presented in ref.~\cite{Bali:2014kia},
has been updated by increasing the statistics and the number
of noisy estimators
to reveal the significant difference visible in figure~\ref{fig:chihom}.}
Altogether, we conclude that the half-half method is insufficient
for our purposes 
and discard it in the following. 

Perturbation theory predicts the dependence
of $\chi_0$ on the lattice spacing, see
equations~(\ref{eq:chiren}) and~(\ref{eq:b1p}).
In figure~\ref{fig:chihom} 
the data are plotted against $\log(a/a_0)$
to verify the expected logarithmic
divergence.
We include the leading $\mathcal{O}(g^2)$ QCD correction to the 
lowest-order QED $\beta$-function coefficient $b_1$.
The renormalization scale $\mu$ is fitted to match the lattice results 
(dashed green line). 
In addition, we multiply the resulting
curve by a rational function that approaches 
unity as $a\rightarrow 0$ (solid yellow error band). This 
band defines the homogeneous magnetic susceptibility
$\chi_0(a)$, as shown for one lattice spacing
in the very left of figure~\ref{fig:lowp}.
The resulting renormalization scale 
reads $\mu=0.123(8)\,\textmd{GeV}$, consistent with our determination 
in ref.~\cite{Bali:2014kia}.

\begin{figure}[ht!]
 \centering
 \includegraphics[width=8cm]{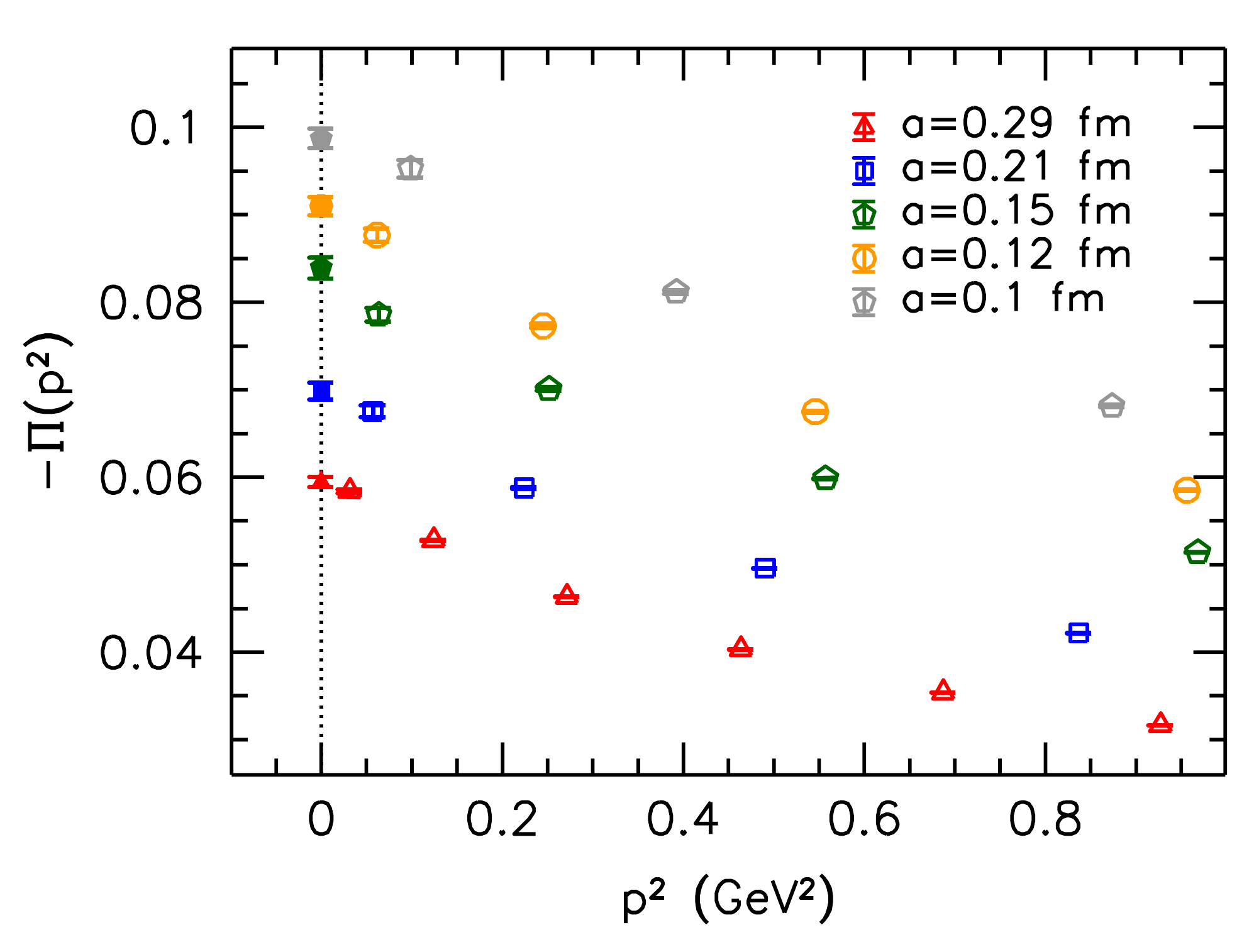}
 \caption{\label{fig:lowp_2}
 Vacuum polarization via magnetic susceptibilities in the low-momentum region. 
The data include both connected and disconnected contributions.
}
\end{figure}

\begin{figure}[ht!]
 \centering
 \includegraphics[width=8cm]{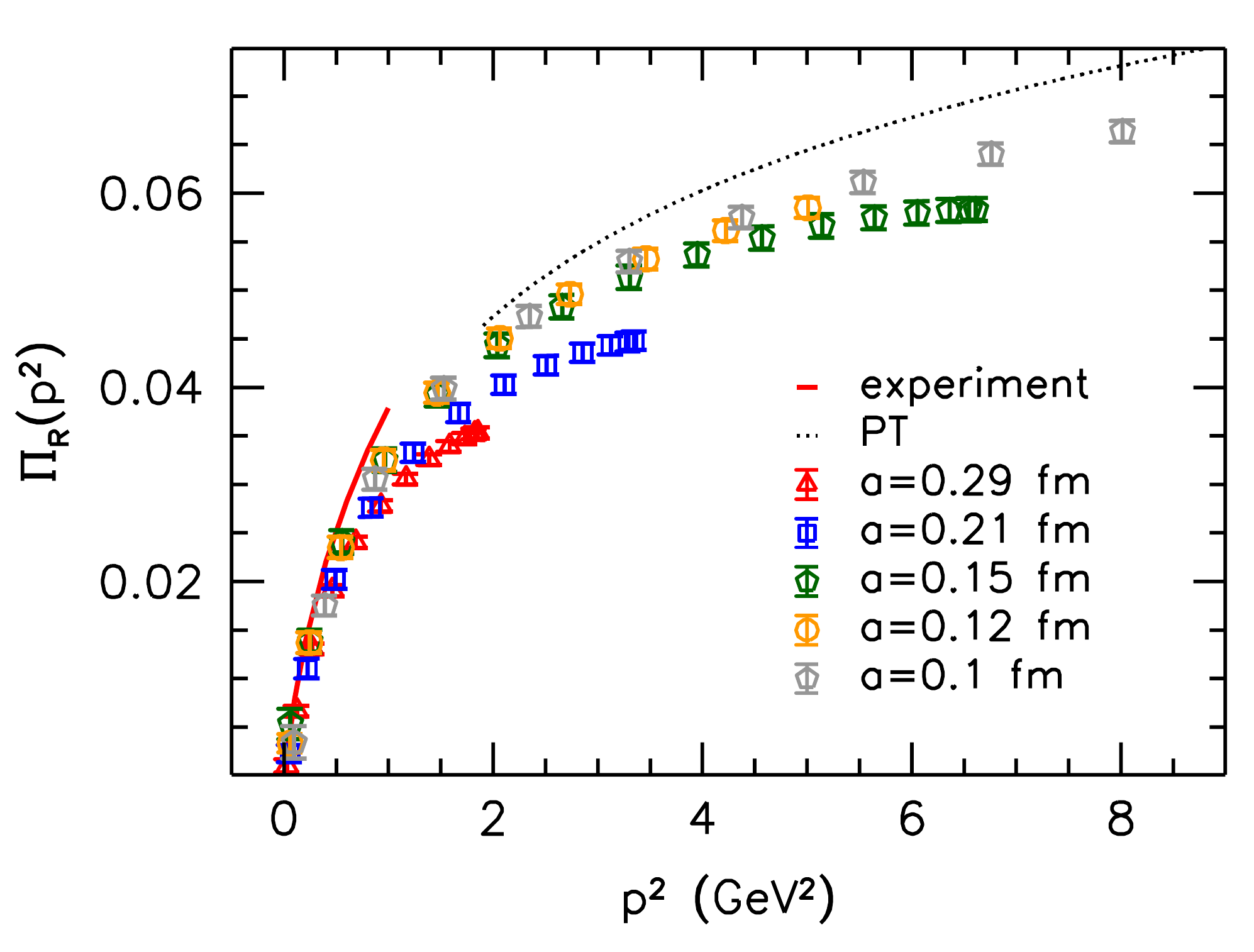}
 \caption{\label{fig:finalres}
 Subtracted vacuum polarization with independent determinations of $\Pi(p^2)$ and 
$\Pi(0)$. The data include both connected and disconnected contributions. 
The solid red line indicates the experimental result (cf.\ figure~\protect\ref{fig:exp})
and the dotted line the three-loop perturbative prediction (see the text).
}
\end{figure}

The $\Pi(p^2)$ results are shown for all five ensembles of table~\ref{tab:1}
in figure~\ref{fig:lowp_2}, where $\Pi(0)=\chi_0$ with the susceptibility
$\chi_0$ determined as detailed above.
Notice that the statistical 
uncertainties (again, both connected and disconnected terms
are taken into account) within our window of lattice spacings
remain at the sub-percent level
for $p^2>0$ and are about one percent for $p=0$. 
Taking into account the statistical errors 
of $\Pi(p^2)$ and of the independently determined $\Pi(0)$,
the renormalized 
vacuum polarization~(\ref{eq:pren})
is plotted in figure~\ref{fig:finalres} for the whole 
momentum region under consideration. For orientation
we also show the three flavor perturbation theory result
for $p^2>2\,\textmd{GeV}^2$, where we truncate the
formulae of refs.~\cite{Baikov:2010je} and~\cite{Baikov:2012zn} 
at $\mathcal{O}(\alpha_s^2)$.
The perturbative curve is
only defined up to an overall constant shift, which we adjust by
matching to a continuum extrapolation around
$p^2=2\GeVt$.
It is clear from the figure that --- as one would expect ---
lattice spacing effects become more prominent towards high momenta.
In addition, the vacuum polarization 
obtained from the experimental $R$-ratio (cf.\ the blue points in figure~\ref{fig:exp}) is included in figure~\ref{fig:finalres}.

Having obtained the renormalized hadronic vacuum polarization,
we can use equations~(\ref{eq:amu}) to (\ref{eq:amu2})
\cite{Lautrup:1971jf,Blum:2002ii} to predict its contribution to the
muon anomalous magnetic moment. 
Choosing a third-order spline interpolation, we obtain values in the range 
$a_\mu^{\mathrm{had,LO}}=(4 \ldots 5)\cdot 10^{-8}$ and an upward trend
towards the continuum limit. This is encouraging as the
$R$-ratio predictions of refs.~\cite{Davier:2010nc} and \cite{Hagiwara:2011af}
for the $N_f=2+1+1$ flavor theory
read $a_\mu^{\mathrm{had,LO}}=6.923 (42)\cdot 10^{-8}$
and $a_\mu^{\mathrm{had,LO}}=6.949 (43)\cdot 10^{-8}$, respectively.
However, given that the present lattices are rather coarse
($0.1\,\textmd{fm}\lesssim a<0.3\,\textmd{fm}$),
we do not yet attempt a full-fledged continuum limit extrapolation.
(Note that at these lattice spacings, the taste splitting of the staggered pion 
multiplet is still sizeable~\cite{Borsanyi:2010cj}.
Thus, large lattice artefacts originating from the heavier pion states are not unexpected, 
since $a_\mu^{\mathrm{had,LO}}$ is highly sensitive to the pseudoscalar masses.)

\subsection{Statistical accuracy and disconnected contributions}

Next, we perform a quantitative comparison between the oscillatory susceptibility method, 
the conventional approach with random wall sources and that with point sources. 
We demonstrate that the statistical error of $\Pi(p^2)$ can be pushed well below that of 
existing studies in the literature -- even with the disconnected terms taken into account.

\begin{figure}
 \centering
 \includegraphics[width=8cm]{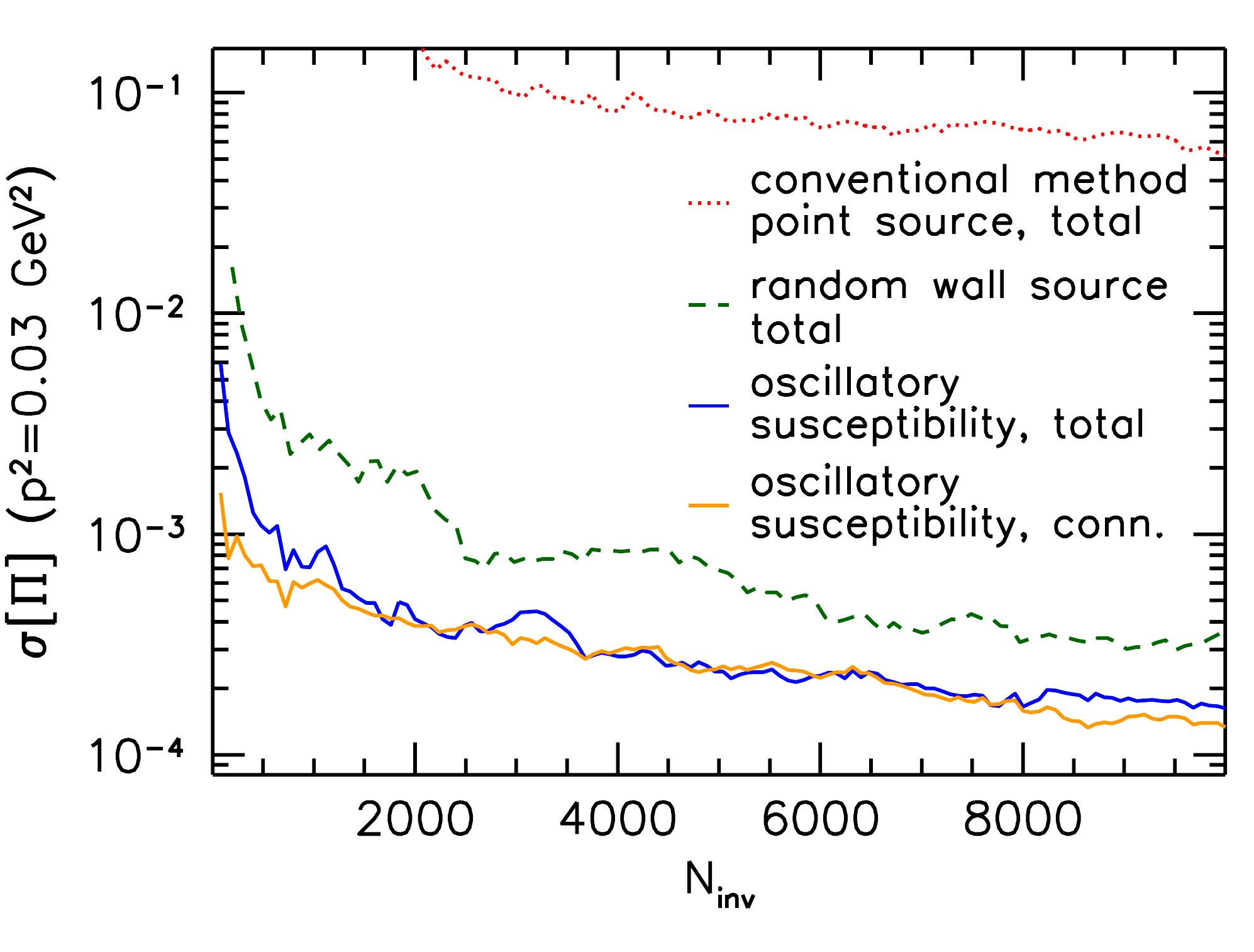}
 \caption{\label{fig:noisy} Statistical error of the 
total (connected plus disconnected)
$\Pi(p^2=0.03\GeVt)$
as a function of the number of inversions. Compared are the results obtained 
from oscillatory susceptibilities, using point sources and
random wall sources. In addition, the error of the connected oscillatory 
susceptibility alone is shown. Note the logarithmic scale.}
\end{figure}

We calculated $\Pi(p^2)$ using all three methods on 120 configurations from the $\beta=3.45$ ensemble 
for a single momentum $p^2=0.03\GeVt$ using an increased number of sources. 
Figure~\ref{fig:noisy} shows the statistical error 
as a function of the number of inversions $N_{\rm inv}$. 
The details of our implementation can be found in appendices~\ref{appB} 
and~\ref{sec:appC}. 
As visible in the figure, the oscillatory susceptibility method allows to save $50-60\%$ of 
the computational effort with respect to the random wall approach. This difference mainly comes
from the disconnected contributions, which can be calculated very accurately via susceptibilities.
In fact, the statistical error in this approach is dominated by the connected 
contribution,\footnote{To see why this is the case, note that the number of estimates increases quadratically 
with $N_{\mathrm{inv}}$ for the 
disconnected terms but only linearly for the connected ones, see the discussion in 
appendix~\ref{appB}. Therefore, 
the error on the latter eventually overtakes that of the former,
before both show the expected asymptotic
$\sigma^2\simeq c_1(1+c_2/N_{\mathrm{inv}})$ fall-off.
The inherent gauge noise $c_1$ can only be reduced by increasing the
number of configurations.}
as is also visible in the 
figure.
As expected, the conventional method with point sources is
not applicable for the determination of the disconnected terms.
Obviously, it is favorable in terms of the total
computer time spent to increase the number of configurations
instead of the number of inversions per configuration. 
We remark that the total number of exact inversions necessary
to achieve a given error can be considerably
reduced by methods like the hopping parameter
expansion~\cite{Thron:1997iy,Michael:1999rs},
truncated eigenmode substitution~\cite{Neff:2001zr,Bali:2005fu,Foley:2005ac},
the truncated solver method~\cite{Collins:2007mh,Bali:2009hu,Blum:2012uh} and,
in the case of Wilson-like fermions, employing spin-explicit
stochastic sources~\cite{Bernardson:1993he,Viehoff:1997wi,Wilcox:1999ab}.

\begin{figure}[ht!]
 \centering
 \includegraphics[width=8cm]{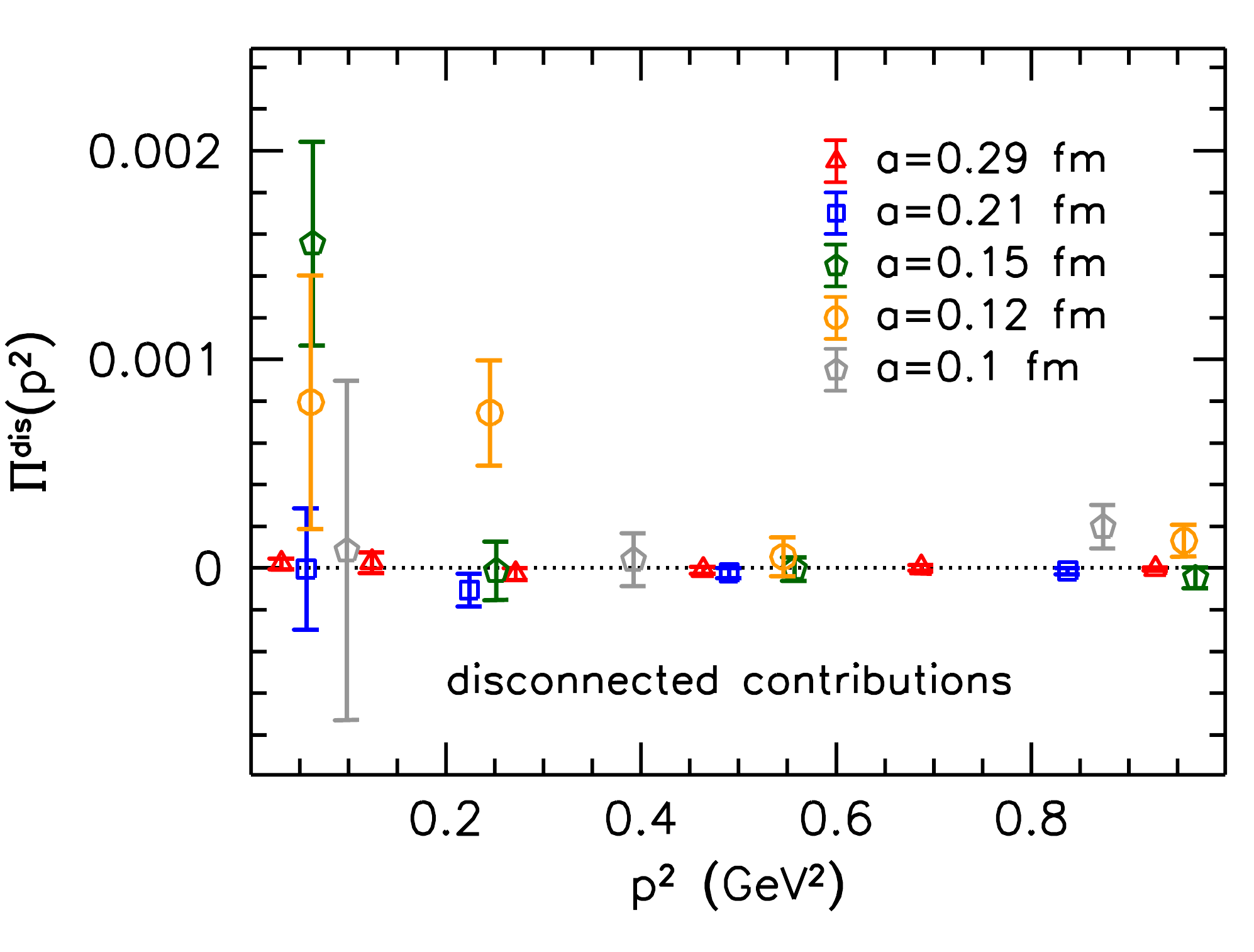}
 \caption{\label{fig:disco}
Disconnected contribution to $\Pi(p^2)$ as a function of $p^2$ for our five
lattice spacings. 
}
\end{figure}

Finally, we discuss the disconnected contribution $\Pi^{\mathrm{dis}}$ in more detail. 
A particular feature of $\Pi^{\mathrm{dis}}$ is that it requires no additive renormalization. 
To see this, note that $\Pi^{\mathrm{dis}}(0)$ vanishes in the perturbative continuum limit, 
since it is of order $g^6(a)$ in the strong coupling~\cite{Baikov:2012zn}, 
which dampens the logarithmic divergence 
and results in $\Pi^{\mathrm{dis}}(0)$ to fall off as $1/\log^2(a)$
for $a\to0$. 
In our three-flavor case the disconnected term even vanishes
identically in perturbation theory
due to $\sum_{f=u,d,s}q_f=0$, once quark masses can be neglected,
i.e.\ $a^{-1}\gg m_s$.
Based on this observation, in figure~\ref{fig:disco} 
we plot the unsubtracted disconnected 
vacuum polarization for all our lattice spacings. 
(The number of inversions was $N_{\mathrm{inv}}=800$ for each momentum,
with the exception of the left-most point.)
Overall, $\Pi^{\mathrm{dis}}$ is consistent with zero,
where the two points that deviate by more than two standard deviations
from this assumption are statistically expected and
no systematic dependence on the lattice spacing or on the volume is apparent.
With the exception of three outliers with large error bars, all central values are below 
$2\cdot10^{-4}$ in magnitude.

Using all available estimators ($N_{\mathrm{inv}}=20\,000$) 
for the $\beta=3.45$ ensemble at $p^2=0.03\,\textmd{GeV}^2$, 
our most accurate 
determinations for the unsubtracted and the subtracted vacuum polarizations read
\be
\begin{split}
p^2=0.03\,\textmd{GeV}^2: \quad\quad \Pi &= -0.058362(117)\,, \\
\Pi^{\mathrm{dis}} &= +0.000021(026)\,, \\
\Pi_{\mathrm{R}} &= +0.002355(198)\,.
\end{split}
\label{eq:numbers}
\ee
Here, $\Pi(p^2)$ and $\Pi^{\mathrm{dis}}(p^2)$ were measured using the oscillatory 
susceptibility method. 
(We highlight again that the error of $\Pi^{\mathrm{dis}}$ is much smaller than that of the total $\Pi$.)
The vacuum polarization at zero momentum was obtained via random wall sources.
Based on the discussion above about the vanishing of $\Pi^{\mathrm{dis}}(0)$ in the 
continuum limit, only the connected part of $\Pi(0)$ is necessary for the subtraction.
The relative error of the so-obtained $\Pi_{\mathrm{R}}$ at this momentum is $8\%$, and 
is dominated by the error of $\Pi(0)$. Clearly, towards higher $p^2$,
where the magnitude of $\Pi(p^2)$ increases, the relative error on $\Pi_{\mathrm{R}}$
rapidly decreases.

\section{Summary}
We developed a new approach to determine the hadronic vacuum polarization $\Pi(p^2)$ on the lattice. 
It is based on calculating magnetic susceptibilities $\chi_p$ with respect to oscillatory background fields 
for $p^2>0$ and a homogeneous background for $p^2=0$. The proof of the equivalence between $\chi_p$ 
and $\Pi(p^2)$ is given in appendix~\ref{sec:appA}. 
The oscillatory susceptibilities are obtained by evaluating the appropriate expectation values 
using noisy estimators, as described in appendix~\ref{appB}. 
Unlike the conventionally used approach, based on position space
current-current correlators, 
which mixes information about all possible lattice momenta, 
the present method enables us to determine the vacuum polarization
with increased precision for individual low momenta. The low momentum region is 
of relevance
for an accurate determination of the 
leading hadronic contribution to the muon anomalous magnetic moment.
In principle, the lattice determination of $\Pi(p^2)-\Pi(0)$ at a selected 
set of low momenta can also be combined with experimental results for the $R$-ratio to increase 
the accuracy of $a_\mu^{\mathrm{had,LO}}$. 

The proposed method not only reduces statistical errors at low momenta but also 
allows for
an independent measurement of $\Pi(0)$, 
instead of having to rely on extrapolations of $\Pi(p^2)$ from $p^2>0$. 
We discussed three different methods to determine the homogeneous susceptibility $\chi_0=\Pi(0)$. 
The most straightforward method, which relies only on simulations at zero magnetic field 
(the so-called half-half method), was found to 
suffer from large finite-volume effects of up to $10\%$ of the full value. Instead, we combined 
existing results on $\chi_0$ from refs.~\cite{Bonati:2013vba,Bali:2014kia} that are 
based on simulations at non-zero background fields.
We also tested stochastic wall sources to obtain
$\Pi(0)$ as the second moment of a momentum projected current-current
correlation function and found that it can compete with the accuracy of the 
homogeneous susceptibility for a sufficiently large number of random sources.
It is interesting to note that $\chi_0$ can also be obtained via stochastic wall sources at 
finite temperatures, giving direct access to the renormalized magnetic susceptibility
$\chi_0(T)-\chi_0(T=0)$ that enters the QCD equation of state at finite magnetic 
fields~\cite{Bali:2013esa,Bonati:2013lca,Levkova:2013qda,Bonati:2013vba,Bali:2013owa,Bali:2014kia}.

\begin{figure}
 \centering
 \includegraphics[width=8cm]{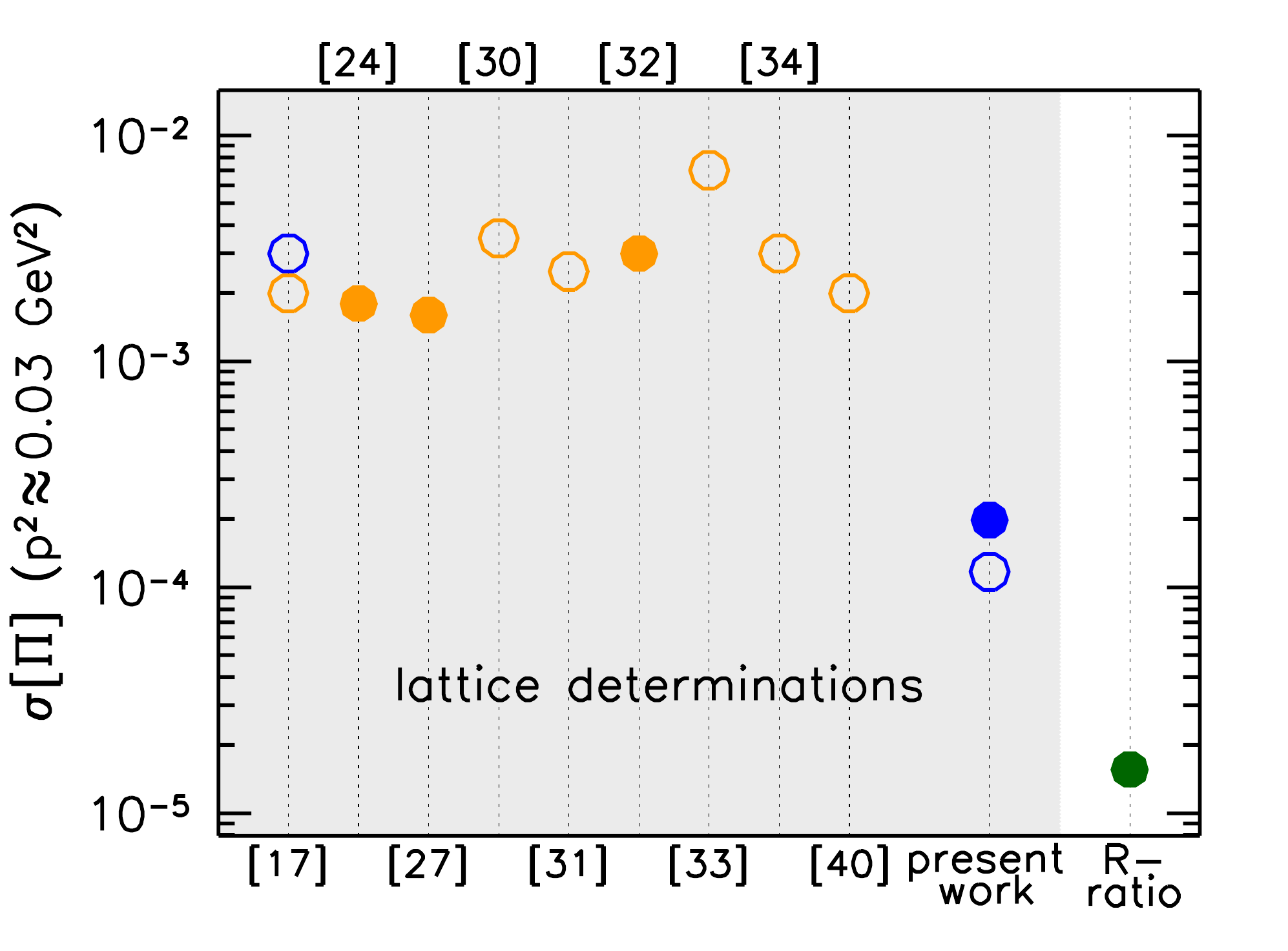}
 \caption{\label{fig:errors} The statistical error of
 the vacuum polarization at low momenta around $p^2=0.03 \GeVt$ for several 
 lattice studies in the literature and for the present work (shaded area).
 Open points denote the error of the unsubtracted $\Pi(p^2)$, while full symbols 
 indicate that of the renormalized $\Pi_{\mathrm{R}}(p^2)$. Studies involving only the 
 connected contribution are indicated in yellow, while those also taking into account the 
 disconnected terms in blue. The determination using the experimental $R$-ratio 
 is also included for comparison (solid green point). 
}
\end{figure}

The method was tested on staggered $N_f=2+1$ flavor ensembles with various lattice spacings.
Already on a few hundred configurations, a statistical accuracy below one percent is achieved
for $\Pi(p^2)$. The disconnected contributions have been included in all cases. 
Figure~\ref{fig:errors} shows an order-of-magnitude comparison
of our statistical accuracy to that of existing 
calculations in the literature, wherever data or figures with error bars are 
available for $\Pi$ at $p^2\approx 0.03 \, \textmd{GeV}^2$~\cite{Burger:2015oya,DellaMorte:2011aa,Bernecker:2011gh,Aubin:2006xv,Boyle:2011hu,Burger:2013jya,Aubin:2012me,Gregory:2013taa,DellaMorte:2014rta,Marinkovic:2015zaa}.
(Note that the approach followed in ref.~\cite{Francis:2013fzp} involves parameterizing 
the lattice data for the zero-momentum projected two-point function $G(t)$ 
of equation~(\ref{eq:get}), 
making a comparison for $\Pi$ difficult.)
We remark that this incomplete comparison does not distinguish between different lattice 
volumes, spacings or pion masses but just serves as a qualitative indicator of 
the accuracy. 
It reveals that our statistical errors, obtained on a comparably
small number of gauge configurations, are by far the smallest within
the lattice 
studies shown in figure~\ref{fig:errors}. 
However, the approach of employing the experimental $R$-ratio 
is still by about an order of magnitude more accurate.
Nevertheless, 
by applying the methods used in this paper to ensembles with substantially higher statistics, 
the desired accuracy may be reached in the near future.

\acknowledgments
This research was funded by the DFG (SFB/TRR 55). The authors acknowledge useful
discussions with Bastian Brandt, Vladimir Braun, Falk Bruckmann,
Pavel Buividovic, Andreas Sch\"afer, K\'{a}lm\'{a}n Szab{\'o}
and in particular
with Bogdan Malaescu and Andreas Hoecker who made available to us
preliminary results on the renormalized hadronic vacuum polarization
function obtained from their analysis of $R$-ratio measurements.
\appendix
\section{Proof of equation~(\ref{eq:mainresult})}
\label{sec:appA}

Below we prove the main result of the paper, equation~(\ref{eq:mainresult}). 
We define the free energy density 
$f=-\log\Z / V_4$, in terms of 
the partition function $\Z$ of the system in a four-dimensional 
volume $V_4$. $\Z$ is 
obtained evaluating the Euclidean functional integral over the 
gluon, quark and antiquark fields $\A_\mu$, $\psi_f$ and $\bar\psi_f$,
\be
\Z = \int \D \A_\mu \prod_{f=1}^{N_f}\D \bar\psi_f \D \psi_f \, e^{-S}\,,\quad
S=\int\dd^4 x \, \L\,,
\label{eq:Partf}
\ee
where the action $S$ 
is the integral of the Lagrange density $\L$. 
Without loss of generality, the magnetic field of equation~(\ref{eq:Bfields}) 
is chosen to point in the third
spatial direction and is generated by a vector potential $\textmd{curl}\,\mathbf A^{\!p} = \mathbf B^{p}$:
\be
A_2^{p} = \int \!\dd x_1 B^{p}\,,\qquad A^{p}_0=A^{p}_1=A^{p}_3=0\,.
\label{eq:gaugeA}
\ee
Here the superscript $p$ indicates the oscillation frequency of the magnetic field, cf.\ equation~(\ref{eq:Bfields}).
The vector potential enters the Lagrange density via minimal coupling:
\be
\begin{split}
\L &= \L_g + \sum_{f=1}^{N_f}\bar\psi_f (\slashed{D}_f^{p}+m_f) \psi_f\,,\\
\slashed{D}_f^{p} &= \gamma_\mu 
\left(\partial_\mu + i \A_\mu + i q_f A^{p}_\mu \right)\,,
\end{split}
\label{eq:Lag}
\ee
where $\L_g$ is the gluonic Lagrangian 
and $m_f$ denote the quark masses. 

In equation~(\ref{eq:gaugeA}) we chose a gauge, in which the photon
vector potential only 
couples to the second component $j_2$ of the electromagnetic current.
Therefore, 
this background probes the $\Pi_{22}(p)$ entry of the vacuum polarization
tensor, where we orient the momentum $p$ to point in the $x$-direction:
$p=(p_1,0,0,0)$. 
In this case, employing equation~(\ref{eq:vacu}), 
the vacuum polarization~(\ref{eq:Piq}) simplifies to
\be
\Pi_{22}(p) =\frac{1}{V_4} \expv{\widetilde{j_2}(p)\widetilde{j_2}(-p)}= -p_1^2 \,\Pi(p^2)\,.
\label{eq:A1}
\ee
For reasons that will become clear in a moment, we consider two
different oscillatory background fields
\be
B^{{\sin},p}(x) = B\sin(px)\,, \qquad B^{{\cos},p}(x)=B\cos(px)\,,
\ee
and denote the corresponding susceptibilities accordingly as $\chi^{\sin}_p$ and 
$\chi^{\cos}_p$. 

Integrating the Lagrange density~(\ref{eq:Lag}) and going to momentum space, 
the magnetic field-dependent part $S_B$ of the action
reads
\be
\begin{split}
S_B(B^{0}) &= B\,\widetilde{j_2}'(0), \\
S_B(B^{{\cos},p}) &= B \left[ \widetilde{j_2}(p)-\widetilde{j_2}(-p)\right]/(2p_1)\,, \\
S_B(B^{{\sin},p}) &= B \left[ \widetilde{j_2}(p)+\widetilde{j_2}(-p)\right]/(2ip_1)\,,
\end{split}
\ee
where the prime denotes differentiation with respect to $p_1$. Inserting these 
expressions into the partition function~(\ref{eq:Partf}) and differentiating twice 
with respect to $eB$ in 
order to obtain the susceptibilities~(\ref{eq:chidef1}) results in
\be
\begin{split}
\chi_0 &= \frac{1}{V_4} \expv{\widetilde{j_2}'(0)\widetilde{j_2}'(0)}\,, \\
\chi^{\cos}_p &= \frac{1}{V_4} \frac{1}{4p_1^2}\expvB{\Big[\widetilde{j_2}(p)-\widetilde{j_2}(-p)\Big]^2},\, \\
\chi^{\sin}_p &= -\frac{1}{V_4} \frac{1}{4p_1^2}\expvB{\Big[\widetilde{j_2}(p)+\widetilde{j_2}(-p)\Big]^2}\,.
\end{split}
\label{eq:A3}
\ee
Note that terms containing the squares of expectation values --- e.g., 
$\langle\widetilde{j_2}'(0)\rangle^2$ for $\chi_0$ ---
vanish due to parity symmetry $B\leftrightarrow -B$ and thus do not appear in 
equation~(\ref{eq:A3}). 

Comparing equations~(\ref{eq:A1}) and~(\ref{eq:A3}) shows that
\be
\chi^{\cos}_p + \chi^{\sin}_p  = \Pi(p^2)\,.
\label{eq:chiPi1}
\ee
In the zero momentum limit, the oscillatory magnetic fields satisfy
\be
\lim_{p\to0}B^{{\cos},p} = B^{0}\,, \qquad
\lim_{p\to0}B^{{\sin},p} = 0\,,\label{eq:Bs}
\ee
which, together with equation~(\ref{eq:chiPi1}), implies for the homogeneous case
\be
\chi_0 = \Pi(0)\,.
\label{eq:chiPi2}
\ee
Furthermore, the $\cos$- and $\sin$-type magnetic fields only differ in a phase and are 
equivalent due to translational invariance. Therefore, the two oscillatory susceptibilities coincide, 
giving:
\be
2\chi_p = \Pi(p^2)\,.
\label{eq:chiPi3}
\ee
Note that the equivalence of $\chi^{\sin}_p$ and $\chi^{\cos}_p$ only holds for
non-zero momenta 
and breaks down at $p=0$. 
In addition, on the periodic lattice the two oscillatory susceptibilities differ at the maximal 
momentum $p_{\max}=\pi/a$ where the $\cos$-type vector potential becomes zero on all lattice sites 
and thus $\chi^{\cos}_{p_{\max}}$ vanishes identically. 
(Still, equation~(\ref{eq:chiPi1}) holds even at this momentum.)

Relations~(\ref{eq:chiPi2}) and~(\ref{eq:chiPi3}) represent the basis of our 
analysis to obtain the vacuum polarization function from magnetic
susceptibilities. We remark that implementing 
equations~(\ref{eq:gaugeA}),~(\ref{eq:A1}) and~(\ref{eq:Bs}) may be thought of as
using $\delta$-sources
in momentum rather than in position space when computing $\Pi(p^2)$.
A similar idea to relate hadronic matrix elements 
to the response to background fields was also discussed in ref.~\cite{Detmold:2004kw}.

\section{Implementation of the susceptibilities}
\label{appB}
In this appendix we present the details of the lattice computation of the 
susceptibilities~(\ref{eq:chidef1}). First of all we have to address
the implications of magnetic flux quantization.

In a finite periodic volume, the magnetic flux $\Phi$
through the perpendicular plane $L_1\times L_2$
(the magnetic field is oriented in the third spatial direction) 
is quantized~\cite{'tHooft:1979uj},
\be
\Phi =\int \!\dd x_1 \,\dd x_2 \, eB = 6\pi N_b\,, \qquad N_b\in \mathbb{Z}\,,
\ee
where we exploited that the smallest electric charge in the system equals $q_d=-e/3$. 
Thus, flux quantization prohibits direct differentiation with respect to the amplitude of the magnetic 
field, unless the flux identically vanishes. For the oscillatory field $B^p(x)$
of equation~(\ref{eq:Bfields}) 
this is indeed the case, making the differentiation with respect to $B$ straightforward. 
For the homogeneous background, the flux is non-zero and, thus, $B$ becomes a discrete 
variable. Various methods to calculate $\chi_0$ are summarized in
refs.~\cite{Bali:2014kia,D'Elia:2015rwa}.

After integrating out the quark fields, the lattice partition function 
becomes an integral over the gluonic links $U_\mu\approx e^{ia\A_\mu}$:
\be
\Z = \int \!\D U_\mu \, e^{-S_g}\! \prod_{f=1}^{N_f}\! \left(\det M_f^{p}\right)^{\!\frac14}\!\!,\quad M_f^{p}=\slashed{D}_f^{p}+m_f\,.
\ee
Here we employed (rooted) staggered quarks to discretize the fermion
matrix $M_f^p$, but the method trivially generalizes to different
discretizations.
The $\mathrm{U}(1)$ vector potential of equations~(\ref{eq:gaugeA}) 
and~(\ref{eq:Lag}) enters $M_f^p$
via the substitution
\be
U_2(x_1) \mapsto U_2(x_1) \cdot e^{ia q_fA_2^p} = 
U_2(x_1)\cdot e^{iaq_fB \frac{\sin(px)}{p_1}},
\label{eq:subs}
\ee
where in the second step we inserted the vector potential for the $\cos$-type magnetic 
field with momentum $p_1$ in the first spatial direction. 
The improvement $p_1\mapsto\hat{p}_1$ is carried out in the denominator of the exponent, 
similarly as in the conventional approach, cf.\ equation~(\ref{eq:latticeWI}).
The derivative with respect to $B$ is then obtained as
\be
\chi_p = \frac{1}{V_4} \expv{\C_{p}^2+\frac{\partial \,{\C}_{p}}{\partial (eB)}}\,,
\ee
where
\be
\C_{p} = 
\frac{1}{4} \sum_f \frac{q_f}{e}\,\tr\left[\left(M_f^{p}\right)^{-1}\dot{M}_f^{p}\,\right]\,,\label{eq:cpp}
\ee
and the dot denotes differentiation with respect to the combination $q_fB$ at $B=0$. 
Having taken the derivative at $B=0$, we can exploit the equality of the up and down quark matrices
$M_u=M_d\equiv M_\ell$ due to the coincident light quark masses. Then, the 
susceptibility reads (suppressing 
the index $p$ and using the electric charge values $q_u/2=-q_d=-q_s=e/3$):
\begin{widetext}
\be
\begin{split}
\chi_p &= \frac{1}{4V_4}\left\langle\frac{5}{9} \tr\left( M_\ell^{-1}\ddot{M_\ell} - M_\ell^{-1}\dot{M_\ell}M_\ell^{-1}\dot{M_\ell}\right) 
+ \frac{1}{9} \tr\left( M_s^{-1}\ddot{M_s} - M_s^{-1}\dot{M_s}M_s^{-1}\dot{M_s}\right) \right\rangle\\
&+ \frac{1}{16V_4} \left\langle\frac{1}{9} \tr\left( M_\ell^{-1}\dot{M_\ell} \right) \tr\left( M_\ell^{-1}\dot{M_\ell} \right) 
+ \frac{1}{9} \tr\left( M_s^{-1}\dot{M_s} \right) \tr\left( M_s^{-1}\dot{M_s} \right)
- \frac{2}{9} \tr\left( M_\ell^{-1}\dot{M_\ell} \right) \tr\left( M_s^{-1}\dot{M_s} \right)
\right\rangle\,,
\end{split}
\label{eq:unimpobs}
\ee
\end{widetext}
where, like in equation~(\ref{eq:cpp}), the pre-factors $1/4$ and $1/16$
are due to the use of rooted staggered fermions.

The first expectation value on the right hand side is the connected
contribution, 
whereas the second one is the disconnected term. 
The traces are measured via a set of noisy estimators
$\xi_j$, $j=1\ldots N_\xi$. 
Taking into account the cyclicity of the trace, the total number of necessary 
inversions is $4N_\xi$ (twice for the light and twice for the strange quark 
matrix). For the calculation of $N_p$ different momenta, some of 
the solutions can be recycled. This results in the total number of
required inversions $N_{\mathrm{inv}}=2N_\xi(1+N_p)$, where the pre-factor
$2$ again is due to $M_s^{-1}\neq M_{\ell}^{-1}$. 
We then have $N_\xi$ independent estimates for the connected 
contribution. Using different stochastic sources for the strange
and for the light quarks, we obtain $N_{\xi}^2$ estimates
of the last disconnected term within equation~(\ref{eq:unimpobs}), 
while for the two non-flavor mixing disconnected terms
we can only exploit
$N_\xi(N_\xi-1)/2$ independent variations.

\section{Implementation of random wall sources}
\label{sec:appC}

Below we specify the details of the calculation of $\Pi(p^2)$ and of $\Pi(0)$ via 
stochastic wall sources. 
In this approach, one calculates the current-current correlator in coordinate space,
and performs the Fourier transformation subsequently. 
Care has to be taken in defining the currents and especially their product at the 
same position. Usually in the literature 
the conserved current is considered and the contact term is
subtracted in order for the 
lattice Ward identity~(\ref{eq:latticeWI}) to hold~\cite{Karsten:1980wd}. 
Another possibility is to consider the product of conserved and
local currents as was done in ref.~\cite{Boyle:2011hu}.

Here we demonstrate how the subtraction of the contact term can be obtained automatically 
if the current is defined using a background $\mathrm{U}(1)$ field $A_\mu$. 
For simplicity, we again consider 
the $\mu=2$ component of the currents and take the distance between the insertions 
to point in the first direction. Then we get
\begin{widetext}
\be
\begin{split}
\expv{j_2(x) j_2(y)} &= \left.\frac{\partial^2 \log\Z}{ \partial A_2(x) \partial A_2(y)}\right|_{A_2=0} \\
&= \frac{1}{4}\expv{\sum_f \left(\frac{q_f}{e}\right)^2 \tr \left[ M_f^{-1} \frac{\partial^2 M_f}{\partial(q_fA_2(x))\partial(q_fA_2(y))} \,\delta_{x,y}-M_f^{-1}  \frac{\partial M_f}{\partial (q_fA_2(x))}
M_f^{-1} \frac{\partial M_f}{\partial (q_fA_2(y))} \right] } \\
&+ \frac{1}{16}\expv{\sum_f \frac{q_f}{e} \,\tr \left[ M_f^{-1} \frac{\partial M_f}{\partial(q_fA_2(x))}\right]}
\expv{\sum_{f'} \frac{q_{f'}}{e} \,\tr \left[ M_{f'}^{-1} \frac{\partial M_{f'}}{\partial(q_{f'}A_2(y))}\right]}.
\end{split}
\ee
Notice that the first term arises due to the fact that 
the background field enters $M_f$ in the 
exponential form $e^{iaq_fA_2}$, and it only contributes if $x=y$. 
Now we define
\be
a^3\!\!\sum_{x_2,x_3,x_4} \frac{\partial M_f}{\partial (q_fA_2(x))}
= \gamma_2 \mathcal{P}_{x_1}\,, \quad\quad
a^3\!\!\sum_{x_2,x_3,x_4} \frac{\partial^2 M_f}{\partial (q_fA_2(x))\partial(q_fA_2(x))}
= \tau_2 \mathcal{P}_{x_1}\,,
\ee
where $\mathcal{P}_{x_1}$ is the projector on the slice of the lattice where 
the first spatial coordinate equals $x_1$. Here, $\gamma_2$ is the staggered discretization of the second 
Dirac matrix and $\tau_2$ its equivalent with the Hermitian conjugate links multiplied 
by minus one,
\be
\begin{array}{c}(\gamma_2)_{xy}\\ (\tau_2)_{xy}\end{array} =
\frac{1}{2} \left[ \eta_2(x) \, U_2(x) \,\delta_{y,x+a\hat{2}}
\begin{array}{c}+\\-\end{array}
\eta_2(x-a\hat{2}) \, U^\dagger_2(x-a\hat{2}) \,\delta_{y,x-a\hat{2}} \right]\,,
\ee
and $\eta_\mu$ denote the staggered phases.
With these definitions we obtain for the two-point function~(\ref{eq:get}) --- 
with the temporal direction replaced by the first spatial direction: 
\be
\begin{split}
G(x_1-y_1) \equiv
\frac{a^6}{L_2L_3L_4}\sum_{x_2,x_3,x_4}\sum_{y_2,y_4,y_4} \expv{j_2(x) j_2(y)} &= 
\frac{1}{4}\expv{\sum_f\left(\frac{q_f}{e}\right)^2 \tr\left[ M_f^{-1} \delta_2 \mathcal{P}_{x_1} 
- M_f^{-1} \gamma_2 \mathcal{P}_{x_1} M_f^{-1} \gamma_2 \mathcal{P}_{y_1} \right] } \\
&+ \frac{1}{16}\expv{\sum_f \frac{q_f}{e} \,\tr\left[ M_f^{-1} \gamma_2 \mathcal{P}_{x_1} \right]}
\expv{\sum_{f'} \frac{q_{f'}}{e} \,\tr\left[ M_{f'}^{-1} \gamma_2 \mathcal{P}_{y_1} \right]}.
\end{split}
\label{eq:G2}
\ee
\end{widetext}
All source positions $y_1$ can be averaged over, keeping the 
distance $x_1-y_1$ fixed, to increase statistics.
Inserting the electric charges and taking into account the degeneracy
of the
light quark masses, this expression can be simplified, in analogy to
equation~(\ref{eq:unimpobs}). For its evaluation we again use noisy estimators 
$\xi_j$ ($j=1\ldots N_\xi$) that are projected using the $\mathcal{P}$ operators. 
One technical issue is the treatment of the second term in the 
connected contribution of equation~(\ref{eq:G2}). Exploiting the $\eta_5$-Hermiticity $M_f^\dagger = \eta_5 M_f \eta_5$ 
of the staggered fermion matrix, the fact 
that $\mathcal{P}^2=\mathcal{P}$ and that the term in question is real, we arrive at
\be
\begin{split}
\xi_j^\dagger \mathcal{P}_{y_1} &\gamma_2  M_f^{-1} \mathcal{P}_{x_1} \gamma_2 M_f^{-1} \mathcal{P}_{y_1} \xi_j \\
&= \left(\mathcal{P}_{x_1} \gamma_2 M_f^{-1} \mathcal{P}_{y_1} \xi_j \right)  \cdot 
\left( \eta_5 M_f^{-1} \eta_5  \gamma_2 \mathcal{P}_{y_1} \xi_j\right)^*\, .
\end{split}
\ee
This demonstrates how this term can be obtained for a fixed source position $y_1$ and 
any sink position $x_1$ using only two inversions.
One of these inversions can also be reused for the calculation of the contact term involving $\tau_2$ 
and for the traces in the disconnected term of equation~(\ref{eq:G2}).
The number of necessary inversions is $N_{\mathrm{inv}}=4N_\xi$.

Putting all this together, the vacuum polarizations at finite and at zero momentum equal
(cf.\ equations~(\ref{eq:vacu}),~(\ref{eq:tmoment}) and~(\ref{eq:latticeWI})),
\be
\begin{split}
\Pi(p^2) &= -\frac{a}{\hat{p}_1^2} \sum_{x_1} e^{i p_1 x_1}\, G(x_1),\\
\Pi(0) &= \frac{a}{2} \sum_{x_1} f(x_1) \,G(x_1),
\end{split}
\ee
where $\hat{p}_1=(2/a)\sin(ap_1/2)$ is the lattice momentum and 
\be
f(x_1) = \begin{cases}
          x_1^2\,, & x_1\le L_1/2\,, \\
	  (L_1-x_1)^2\,, & \textmd{otherwise}
         \end{cases}
\ee
is a quadratic function
consistent with the boundary conditions for a periodic lattice 
with linear size $L_1$. 
We mention that the separation $x-y$ of equation~(\ref{eq:G2}) is usually chosen to lie in the 
temporal direction, as indicated in equation~(\ref{eq:tmoment}). In 
our setup it points in the first spatial direction to make the connection 
to the magnetic susceptibilities -- 
involving $x_1$-dependent phases, cf.\ equation~(\ref{eq:subs}) -- 
more transparent.
\bibliography{Adler}
\end{document}